\documentclass[aps,pre,twocolumn,showpacs,showkeys]{revtex4}

\usepackage{graphicx}
\usepackage{dcolumn}
\usepackage{bm}
\usepackage{amssymb}
\usepackage{amsmath}
\usepackage{array}
\usepackage{rotating}
\usepackage{color}

\begin{document}

\preprint{IRE Kharkov}

\title{Analysis and optimization of a free-electron laser \protect\\ with an irregular waveguide}

\author{V.A. Goryashko}
\email{vitgor@ire.kharkov.ua}
 \altaffiliation{Institute for Radiophysics and Electronics of NAS of Ukraine, \\
 12  Acad. Proskura Street, Kharkiv, 61085, Ukraine.}

\date{\today}

\begin{abstract}

Using a time-dependent approach the analysis and optimization of a
planar FEL-amplifier with an axial magnetic field and an irregular
waveguide is performed. By applying methods of nonlinear dynamics
three-dimensional equations of motion and the excitation equation
are partly integrated in an analytical way. As a result, a
self-consistent reduced model of the FEL is built in special phase
space. The reduced model is the generalization of the
Colson-Bonifacio model and takes into account the intricate dynamics
of electrons in the pump magnetic field and the intramode scattering
in the irregular waveguide. The reduced model and concepts of
evolutionary computation are used to find optimal waveguide
profiles. The numerical simulation of the original non-simplified
model is performed to check the effectiveness of found optimal
profiles. The FEL parameters are chosen to be close to the
parameters of the experiment (\textit{S. Cheng et al. IEEE Trans.
Plasma Sci. 1996, vol. 24, p. 750}), in which a sheet electron beam
with the moderate thickness interacts with the TE$_{01}$ mode of a
rectangular waveguide. The results strongly indicate that one can
improve the efficiency by a factor of five or six if the FEL
operates in the magnetoresonance regime and if the irregular
waveguide with the optimized profile is used.
\end{abstract}

  \maketitle

\section{Introduction}

The recent progress in the theory and experiment of free-electron
lasers (FELs) and gyrotrons~\cite{Ginzburg_2000,Temkin_2001}  with
Bragg cavities is strongly indicative that the application of novel
electrodynamical structures provides the opportunity to realize
unique properties of FELs to a large measure. For example, Bragg
cavities, which are periodic arrays of varying dielectric or
metallic structures, stimulate interest in traditional microwave
applications because they can be built oversized (quasioptical) and,
therefore, employed at higher frequencies and higher power. At the
same time the investigation of traveling wave tubes
(TWT)~\cite{Kuraev_2007} shows that the TWT efficiency based on a
regular (along the interaction region) electrodynamical structure is
far from its possible maximal value. In fact, the difference between
the cold phase velocity and the average velocity of the beam is the
control parameter of the beam-wave interaction. By changing this
parameter along the interaction region one can control the beam
bunching and the energy transfer between bunches and microwaves. The
local variation of the cold phase velocity along the region depends
on the local variation in the waveguide profile. Then, in an effort
to control the beam-wave interaction and improve the efficiency one
should use irregular electrodynamic structures. Specifically, the
combination of Bragg reflectors~\cite{Bratman_1983} and the section
of an irregular waveguide seems to be a promising electrodynamic
structure for a high-efficiency powerful FEL. The analysis of a
planar FEL-amplifier with an axial magnetic field and an irregular
waveguide is the topic of the present paper. I focus my attention on
this FEL configuration because it is well known that through the use
of the axial magnetic field one can substantially improve the
efficiency, as the cyclotron frequency tends to the undulator
frequency~\cite{Sprangle_1978}.

It is worth noting that in vacuum electronic sources of coherent
radiation the electron beams are far from the statistical
equilibrium and during their interaction with radiation they remain
sufficiently nonequilibrium because of the large free
length~\cite{VanSoln_1973,Davidson_book}. Thus, the efficiency of
the transfer of the electrons' kinetic energy into radiation,
basically, may be close to 100\% (the klystron or traveling wave
tube are the examples of high-efficiency devices) and the challenge
is to optimize the beam-wave interaction by controlling the most
important parameters. There are several ways to improve the FEL
efficiency: optimization of electron beam characteristics (for
example, development of beams with the optimal distribution of the
axial velocity across a beam when the effect of beam finite
thickness is relevant), tapering of the undulator or the axial
magnetic field, profiling of waveguide/resonantor walls. In
particular, the effectiveness and reliability of the undulator
tapering were demonstrated
theoretically~\cite{Kroll_1981,Freund_1984_lett} and confirmed
experimentally to a great advantage. In the
experiment~\cite{Fawley_1986} the saturated power of 180 MW (6\%
efficiency) has been increased to 1.0 GW (34\% efficiency) by
optimizing the wiggler profile in such a way that the beam-wave
resonance condition remains fulfilled for many electrons, as the
electrons lose their energy. The numerical simulation mentioned
in~\cite{Fawley_1986} indicates that one can trap about 75\% of the
electron beam and reduce its energy by about 45\%. A high
effectiveness of the undulator tapering was also demonstrated for a
FEL with an axial guide magnetic field~\cite{Freund_1986_v33} (the
maximum efficiency was increased by almost 700\% as compared to the
untapered configuration). At the same time there exist cases where
the convenient undulator profiling cannot be used or ensure desired
enhancement. In particular, if the waveguide backward mode is used
as an electromagnetic undulator, then, clearly, one has to change
electrodynamic structure characteristics to control electromagnetic
undulator parameters. The optimization of the electromagnetic
structure also seems to be more efficient than the undulator
profiling if the effect of the electron beam finite thickness is
relevant. Indeed, given the FEL amplifier operates with Group~II
orbit parameters, a negative masslike effect
occurs~\cite{FreundAntonsen_1995} in which the electrons are axially
accelerated, as they lose energy to the wave. Hence, the electrons
must be decelerated to maintain the beam-wave resonance. This is
accomplished by an upward taper in the undulator
field~\cite{Freund_1986_v33}. At the same time, the electrons with
different initial transverse positions are exposed to the action of
different magnitudes of the pump magnetic field. As a result, the
average velocity of the electron depends on its initial transverse
position, and different beam layers have different detunings with
the wave because the average velocity of the electron governs the
initial detuning. If the undulator field is tapered `up', then the
detuning of external layers of a thick beam increases and the
contribution of these layer to the total efficiency shows a certain
decrease. In the present paper I demonstrate that one can
effectively suppress layering and the saturation efficiency effect
by using the optimized profiled waveguide. I believe this technique
to be useful for the development of powerful thick-beam FELs (for
example the FEM experiments~\cite{Arzhannikov_2009} on generation
and transport of two intense beams have been performed of late:
0.8~MeV electron energy, current densities of up to 1.5~kA/cm$^2$,
0.4 $\times$ 7.0~cm$^2$ beam cross sections).

This paper is structured as follows: in Sec.~II the problem
statement for the planar FEL-amplifier with the axial magnetic field
and the irregular waveguide is defined. The integrals of motion for
a test electron in the pump magnetic field are constructed in
Sec.~III. With these integrals and the method of nonlinear resonance
the FEL reduced model is derived in Sec. IV. In the following
section the principle of the beam-wave control are considered and a
practical example of the optimized FEL is given. The obtained
results are discussed in Summary, and, finally, the wave interaction
in the irregular waveguide is examined in Appendices A and B.

\section{The theoretical model}\label{section:model}

\begin{figure}[!b]
\centering %
\scalebox{1.0}{\includegraphics[222,636][447,757]{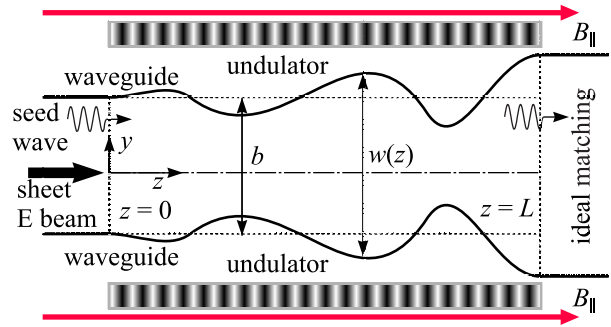}}
\caption{\label{irregular_waveguide} Sketch of the FEL in the $x=0$
cross section.}
\end{figure}

Let a sheet relativistic electron beam be injected into an irregular
waveguide located in the external pump magnetic field that consists
of the magnetic field of a linearly polarized (planar) undulator and
a uniform axial magnetic field (see Fig.~\ref{irregular_waveguide}).
The pump magnetic field is given by the vector-potential:
\begin{equation}\label{pump_field}
{A}_x^p(\vec{r}) = (B_u/k_u) \cosh(k_u y)\cos(k_u z) + B_\| y.
\end{equation}
Here $B_\|$ is the uniform axial guide field, $B_{u}$ is the
magnitude of the planar undulator field~\cite{Destler_1994}, $k_u =
2\pi/\lambda_u$ and $\lambda_u$ are the wave number and the period
of the undulator, respectively. In numerical simulations we model
the injection of the electron beam into the interaction region by
allowing the undulator amplitude to increase adiabatically from zero
to a constant level over $N_u$ undulator periods~\cite{Freund_1987}.
The unmodulated electron beam enters the interaction region, $z\in
[0,L]$, with average longitudinal velocity $V_{\|}$. The irregular
waveguide boundaries are set by expressions: $x=\pm a/2$ and $y=\pm
w(z)/2$ ($a \gg w$), where $w(z)$ describes the varying distance
between two wide walls of the waveguide, and $w'(0)=w'(L)=0$. Let
the FEL-amplifier be seeded by the TE$_{01}$ mode, which is resonant
(synchronous) with the electron beam, the mode frequency and the
amplitude at the input into the interaction region ($z=0$) equaling
$\omega$ and $V_0$, respectively. We consider that the interaction
region is ideally matched to the regular output waveguide at the
section $z=L$.

Since the narrow walls are profiled, $y=\pm w(z)/2$, one can apply
the local Fourier-series expansion~\cite{Tarasov_Makarov} over $y$
to derive the coupled set of equations governing the evolution of
TE$_{nm}$ and TM$_{nm}$ modes (subscripts $n$ and $m$ correspond to
the field variation along the wide and narrow walls, respectively).
Modes with odd TE$_{n,\mathrm{odd}}$, TM$_{n,\mathrm{odd}}$ and even
TE$_{n,\mathrm{even}}$, TM$_{n,\mathrm{even}}$ variations are not
coupled because of the waveguide symmetry with respect to
$xz$-plane. We will hold that $\lambda>2w(z)/3$, and the TE$_{0m}$
modes for $m>3$ are then evanescent and the scattering of the seed
TE$_{01}$ mode to those modes will be neglected. We also ignore the
electron beam mode generation. Under this assumptions the evolution
of the signal TE$_{01}$ mode is governed by the $x$-component of
vector-potential $A_x^s$:
\begin{equation}\label{microwave}
  A_x^s(\vec r, t) = \mathrm{Re}\Bigl\{V (z,t)\sqrt{\frac{b}{w(z)}}
        \cos\Bigl(\frac{\pi y}{w(z)}\Bigr) \,e^{-i\omega t}\Bigr\}.
\end{equation}
Here $V(z,t)$ is the slow-in-time amplitude satisfying the equation
(see~\ref{appendix1})
\begin{multline}\label{nonregular_amplitude_time}
  \Bigl\{\frac{\partial^2}{\partial z^2} + k_z^2 + \frac{2ik_z}{v_{gr}}\frac{\partial}{\partial t}\Bigl\}V(z,t)
        =  -\frac{8\omega}{Sc} \sqrt{\frac{b}{w(z)}}\int_{-a/2}^{a/2}  dx \times
  \\
                  \int_{-w(z)/2}^{w(z)/2} dy \int_{t-\pi/\omega}^{t+\pi/\omega} dt'\,
                  j_x(\vec r,t')  \cos\Bigl(\frac{\pi y}{w(z)}\Bigr) e^{i\omega t'},
\end{multline}
where
\begin{equation*}
  k_z^2(z)= \frac{\omega^2}{c^2} - \Bigl(\frac{\pi}{w(z)}\Bigr)^2
       - \Bigl(\frac{w'(z)}{2w(z)}\Bigr)^2\Bigl[1+\frac{\pi^2}{3}\Bigr]
\end{equation*}
is the wave number squared,  $v_{gr}(z,\omega) =
(dk_z/d\omega)^{-1}$ is the group velocity, $c$ is the speed of
light and $S=a\times b$. The boundary conditions read
\begin{equation}\label{boundary_condit_time}
\begin{split}
  & \Bigl(\frac{\partial V}{\partial z} + ik_z V - \frac{1}{v_{gr}}\frac{\partial V}{\partial t} \Bigr)\Bigr|_{z=0} = 2ik_z V_0,
  \\
  & \Bigl(\frac{\partial V}{\partial z} - ik_z V + \frac{1}{v_{gr}}\frac{\partial V}{\partial t} \Bigr)\Bigr|_{z=L} = 0.
\end{split}
\end{equation}

The microscopic current density is given by the following sum over
electron trajectories,~\cite{Freund_1987}:
\begin{multline}\label{charge_density}
  \vec{j}(\vec{r},t) = \frac{I_0}{S_b} \!\!\int\limits_{-X_b/2}^{X_b/2}\!\! dx_0
                        \!\!\int\limits_{-Y_b/2}^{Y_b/2}\!\! dy_0
                        \!\!\int\limits_{t - L/V_\|}^{t}\!\! dt_e
   \frac{\vec p (z;\vec{r\;}_{\perp 0},t_e)}{p_z(z;\vec{r\;}_{\perp 0},t_e)} \times
 \\
 \delta[\vec{r}_\perp - \vec{r}_\perp(z;\vec{r\;}_{\perp 0},t_e)]
 \delta[t-t(z;\vec{r\;}_{\perp 0},t_e)],
\end{multline}
where $I_0$ is the beam current at the input into the interaction
region; $S_b=X_b Y_b$ is the cross sectional area of the beam; $\vec
p (z;\vec{r}_{\perp 0},t_e)$ and $\vec r_\perp(z;\vec {r}_{\perp
0},t_e)$ are the mechanical momentum and the transverse coordinate,
respectively; $t(z;\vec{r}_{\perp 0},t_e)$ is the arrival time of an
electron at the position $z$; $t_e$ and $\vec{r}_{\perp
0}=\vec{r}_{\perp 0}(x_0,y_0)$ are the entrance time and the
transverse coordinates, which the electron has at the input of the
interaction region. The sheet electron beam is lying from
$x_0=-X_b/2$ to $x_0=X_b/2$ and from $y_0=-Y_b/2$ to $y_0=Y_b/2$ in
the $x$ and $y$ directions, respectively. Since the relativistic
electron-wave interaction is being studied, the nonradiated fields
(space-charge fields) are supposed to be negligible. The
relativistic effects result in that the force $f_m$ caused by the
nonradiated magnetic field partially suppresses the defocusing
action of the transversal part of the force $f_e^{(pot)}$ caused by
the potential part of the nonradiated electric field, the axial
component of $f_e^{(pot)}$ being partially compensated by the force
$f_e^{(rot)}$ caused by the rotational part of the nonradiated
electric field (see~\cite{We} for details).

The motion of a typical electron within the electron beam can be
described by the relativistic Hamiltonian
\begin{equation}\label{Hamiltonian_full}
  \mathcal{H} = \sqrt{m_e^2 c^4 + \bigl(c\vec P - e\vec A^p - e\vec A^s \bigr)^2}
  = m_e\gamma c^2.
\end{equation}
Here $e$ and $m_e$ are the electron charge and rest mass,
respectively; the canonical momentum $\vec P$ is related to the
mechanical momentum $\vec p$ by $\vec P =\vec p + (e/c)(\vec
A^p+\vec A^s)$. The initial conditions for the mechanical momentum
and coordinates read:
\begin{equation}
\begin{aligned}
  & p_x|_{t=t_e}=p_y|_{t=t_e}=0, \quad p_z|_{t=t_e}=\mathcal{E} V_\|/c^2,
  \\
  & x|_{t=t_e} = x_0, \quad y|_{t=t_e} = y_0, \quad z|_{t=t_e}=0,
\end{aligned}
\end{equation}
where $\mathcal{E}$ is the initial energy of the electron entering
the interaction region at the time $t_e$. The excitation equation
\eqref{nonregular_amplitude_time} along with the expression for the
current density~\eqref{charge_density} and the equations of motion
generated by the Hamiltonian~\eqref{Hamiltonian_full} describe the
electron-wave interaction in the studied FEL in a self-consistent
way. In the self-consistent models the main mathematical
obstructions are due to the nonlinear character of the equations of
motion, and in order to perform the analytical treatment of the FEL
we first have to integrate the equations of motion generated by the
Hamiltonian~\eqref{Hamiltonian_full}. For the purpose of subsequent
analysis, it is convenient to rewrite the
Eq.~\eqref{Hamiltonian_full} as
\begin{equation}\label{Hamiltonian_perturbed}
  \mathcal{H}(\vec r, \vec P, t) = \sqrt{ \bar{\mathcal{H}}^2(\vec r, \vec P) + W(\vec r, \vec P, t)},
\end{equation}
where $\bar{\mathcal{H}}$ describes the dynamics of a typical
electron in the pump magnetic field
\begin{multline}\label{Hamiltonian_upperturbed}
  \bar{\mathcal{H}} = \Bigl[\bigl\{c P_x - e[(B_u/k_u) \cosh(k_u y)\cos(k_u z) + B_\| y] \bigr\}^2+
  \\
                     (c P_y)^2+(c P_z)^2 + m_e^2 c^4 \Bigr]^{1/2} = m_e\gamma_0 c^2,
\end{multline}
and the ponderomotive perturbation reads
\begin{equation}
  W = - 2e(c P_x - e A_x^p) A_x^s + (e A_x^s)^2.
\end{equation}

We start our analysis with the integration of the equation of motion
generated by the unperturbed Hamiltonian
\eqref{Hamiltonian_upperturbed}. The above integration is the
nontrivial problem because the nonlinear dynamical
system~\eqref{Hamiltonian_upperturbed} is not globally
integrable~\cite{Buzzi_1993} and exhibits chaotic behavior if the
absolute value of the difference between the normal undulator and
normal cyclotron frequencies is less than the betatron
frequency~\cite{We}. The dynamics of electrons in an ideal undulator
and an axial magnetic field was studied in details in Ref.~\cite{We}
using Lindshtedt's perturbation method in configuration space. This
allowed one to build a linear microwave theory and analyze the
electron-wave interaction in the magnetoresonance regime. However,
to build the nonlinear theory we have to study the behaviour of a
test electron in the pump magnetic field in a more specific way. In
the next section we build the approximate solution
for~\eqref{Hamiltonian_upperturbed} by means of the superconvergence
method in action-angle variable space and derives the explicit
analytical expressions for the region location of chaotic dynamics
in the parameter space.

\section{Dynamics of electrons in the pump magnetic field}
\label{section:Dynamics}
\subsection{Action-angle variables formulation}

In order to apply the perturbation method to the nonlinear system
with the Hamiltonian \eqref{Hamiltonian_upperturbed} the latter is
divided into a nonperturbed (integrable) part that corresponds to
the electron motion in the axial homogeneous magnetic field and a
small perturbation caused by the undulator magnetic field. Based on
the nonperturbed system (the undulator field is absent $B_u=0$) we
can introduce the action-angle variables:
\begin{equation}\label{coordiantes_chahge}
\begin{aligned}
 P_y & = -\frac{\sqrt{2\omega _{c}\mathcal{E}I_c}}{c}\sin\vartheta _c,
     \qquad  & P_z & = k_u I_u,
  \\
  y & = c\sqrt{\frac{2I_c}{\omega _{c}\mathcal{E}}}\cos\vartheta _c -\frac{c^2P_x}{\omega_{c}\mathcal{E}},
      & \quad z & = \frac{\vartheta_u}{k_u};
\end{aligned}
\end{equation}
the initial conditions take the following form:
\begin{equation}\label{in_con_1}
\begin{aligned}
   I_u\bigl|_{t=t_e} & = \beta_\|\mathcal{E}/(k_u c),  \quad  & \vartheta_u\bigl|_{t=t_e} & = 0,
   \\
   I_c\bigl|_{t=t_e} & = \varepsilon^2 \beta_\|^2 \mathcal{E} \cosh^2[k_u y_0]/(2\omega_c),  & \vartheta_c\bigl|_{t=t_e}& = \pi,
\end{aligned}
\end{equation}
where $\beta_\| = V_\|/c$, $\omega_c=|e|cB_\|/\mathcal{E}$ and
$\omega_u=k_u V_\|$ are the partial cyclotron and undulator
frequencies, respectively;
$\varepsilon=\sqrt{2}\omega_\beta/\omega_u$ is the dimensionless
perturbation para\-me\-ter,
$\omega_\beta=|e|cB_\|/(\sqrt{2}\mathcal{E})$ is the betatron
frequency.

The electron motion described by the Hamiltonian
Eq.~\eqref{Hamiltonian_upperturbed} is characterized by two degrees
of freedom, namely, by undulator (subscript $u$) and cyclotron
(subscript $c$) degrees of freedom. The transverse inhomogeneous of
the realistic undulator field does not lead to the appearance of the
additional betatron degree of freedom, but only modifies the
undulator and cyclotron motion. For instance, if
$\omega_\beta,\omega_c\ll \omega_u$ then the undulator and cyclotron
frequencies of coupled nonlinear oscillations are $\Omega_u =
\{\omega_u^2 -3 \omega_\beta^2 \cosh^2[k_u y_0]\}^{1/2}$ and
$\Omega_c = \{\omega_c^2 + \omega_\beta^2 \cosh^2[k_u y_0]\}^{1/2}$.
Note that in purely undulator field ($B_\|=0$) the undulator and
betatron degrees of freedom become split and the dimension of the
dynamical system is also equal to two~\cite{ChenDavidson1990}.
Using~\eqref{coordiantes_chahge} and some algebra we may
rewrite~\eqref{Hamiltonian_upperturbed} as
\begin{equation}
    \bar{\mathcal{H}} = \sqrt{ m_e^2c^4 + (ck_uI_u)^2 + 2\omega_c\mathcal{E}I_c + \varepsilon\, V(\vec I,\vec \vartheta)}.
\end{equation}
Here $\varepsilon\, V(\vec I,\vec \vartheta)$ is the nonintegrable
undulator perturbation
\begin{multline}\label{perturbation}
  \varepsilon\,V(\vec I, \vec\vartheta) =
                  \sum\limits_{n=0}^{2}\sum\limits_{m=0}^{\infty}V_{n,m}(I_c)
                  \bigl(\cos[n \vartheta_u + m\vartheta_c]+
\\
                  \cos[n \vartheta_u - m\vartheta_c]\bigr),
\end{multline}
where for odd values of $m$ the coefficient $V_{n,m}$ is
\begin{equation*}
\begin{split}
    V_{0,m} &= V_{2,m} = -\varepsilon^2\beta_\|^2\mathcal{E}^2 \sinh[2A]I_m[2B]/4,
\\
    V_{1,m} &= \varepsilon\beta_{\|}\mathcal{E}\sqrt{2\omega_c\mathcal{E}I_c}\cosh[A](I_{m+1}[B] + I_{m-1}[B]),
\end{split}
\end{equation*}
and for even values of $m$ the coefficient $V_{n,m}$ is
\begin{equation*}
\begin{split}
    & V_{0,m} = \varepsilon^2\beta_\|^2\mathcal{E}^2(\delta_{m,0}+\cosh[2A]I_{m}[2B](2-\delta_{m,0}))/8,
\\
   & V_{2,m} = V_{0,m}, \quad V_{1,m} =-\varepsilon\beta_{\|}\mathcal{E}\sqrt{2\omega_c\mathcal{E}I_c} \sinh[A]\times
\\
    &\hspace{3.5cm}(I_{m+1}[B] + I_{m-1}[B](1-\delta_{m,0})),
\end{split}
\end{equation*}
Here,  $A =
-(k_uy_0+(\sqrt{2}\omega_\beta/\omega_c)\cosh[k_uy_0])=\mathrm{const}$,
$B=ck_u\sqrt{2I_c/(\omega_c\mathcal{E})}$, $\delta_{m,n}$ is the
Kronecker delta, $I_n(x)$ is the modified Bessel function of the
first kind of order $n$. The equations of motion read:
\begin{equation}\label{eq_I_theta}
\begin{split}
  \biggl\{ & {\dot I_u \atop \dot I_c} \biggr\}  = \sum\limits_{n=0}^{2}\sum\limits_{m=0}^{\infty}
          \biggl\{ {n \atop m} \biggr\}
         \frac{V_{n,m}(I_c)}{2 \mathcal{E}}\bigl(\sin[n \vartheta_u+ m \vartheta_c]\pm
\\
                                           &\hspace{4.7cm}\sin[n \vartheta_u- m \vartheta_c]\bigr),
\\
  & \dot\vartheta_u = c^2 k_u^2 I_u/\mathcal{E},
\\
 & \dot\vartheta_c  = \omega_c + \sum\limits_{n=0}^{2}\sum\limits_{m=0}^{\infty}\frac{\partial V_{n,m}(I_c)}{2\mathcal{E}\; \partial I_c}
                                   \bigl(\cos[n \vartheta_u + m\vartheta_c]+
\\
                                    &\hspace{4.7cm}\cos[n \vartheta_u - m\vartheta_c]\bigr).
\end{split}
\end{equation}
The equation set \eqref{eq_I_theta} has a lot of internal nonlinear
resonances ($\vartheta_u\approx \pm m \vartheta_c$,
$2\vartheta_u\approx \pm m \vartheta_c$) between the undulator and
cyclotron degrees of freedom. Actually, the successive iterations
give in the zero approximation $\vec I^{(0)} = \vec I\bigl|_{t=t_e}$
and $\vartheta_u^{(0)}=\omega_u(t-t_e)$, $\vartheta_c^{(0)}=\pi +
\omega_c(t-t_e)$. One can check that the first approximation leads
to $\vec \vartheta^{(1)},\vec I^{(1)}\propto e^{i(n\omega_u\pm
m\omega_c)t}/(n\omega_u\pm m\omega_c)$. As a result, the close is
$n\omega_u\pm m\omega_c$ to zero, the more perturbed dynamics is.
Applying the nonlinear resonance technique to~\eqref{eq_I_theta} and
analyzing each internal nonlinear resonance separately we can show
that $I_c\propto \sqrt{\varepsilon}$ in the vicinity of the
resonance. Using this estimation we can compare the levels of the
dominance of different resonances ($\varepsilon\ll1$,
$k=1,2,3\ldots$):
\begin{equation}\label{resonances}
\begin{aligned}
  \mathrm{if\;}\vartheta_u &\approx (2k-1)\vartheta_c   &\mathrm{ then\quad }&
    \varepsilon\,V\propto \varepsilon^{k/2+3/4}\cosh[A],
\\
  \mathrm{if\;}\vartheta_u &\approx 2k\vartheta_c   &\mathrm{ then\quad }&
    \varepsilon\,V\propto \varepsilon^{k/2+1}\sinh[A],
\\
  \mathrm{if\;}2\vartheta_u &\approx (2k-1)\vartheta_c   &\mathrm{ then\quad }&
    \varepsilon\,V\propto \varepsilon^{k/2+7/4}\sinh[2A],
\\
  \mathrm{if\;}2\vartheta_u &\approx 2k\vartheta_c   &\mathrm{ then\quad }&
    \varepsilon\,V\propto \varepsilon^{k/2+2}\cosh[2A].
\end{aligned}
\end{equation}
It will be further shown that if $\varepsilon$ and $y_0$  exceed
some thresholds, then there exist regions of chaotic dynamics of the
test electron in the phase space. These regions of the phase space
correspond to the regions of the nonlinear resonances between
degrees of freedom. The most important resonances, in the vicinity
of which the onset of the chaos can occur, are $\vartheta_u \approx
\vartheta_c$, $\vartheta_u \approx 2\vartheta_c$ and $\vartheta_u
\approx 3\vartheta_c$ (see Fig.~\ref{k&s_Y_005_e_02_NoneAdiabat},
\ref{k&s_Y_005_e_02_Adiabat}).

\subsection{Superconvergent method}

An efficient method for analytical treatment of Hamiltonian systems
is the application of canonical transformations to a
Hamiltonian~\cite{Goldstein_1975}, \cite{Arnold_1979},
\cite{Chirikov_1979}. Then we seek for canonical transformations to
new dynamical variables such that a new Hamiltonian
$\bar{\mathcal{H}}'$ is a function of action variables only.
Therefore, new actions become integrals of motion. According to the
superconvergence method~\cite{Chirikov_1979} we choose successive
canonical transformation $(\vec I,\vec \vartheta)\rightarrow$ $(\vec
I_1,\vec \vartheta_1)\rightarrow$ $(\vec I_2,\vec \vartheta_2)
\rightarrow$ $\ldots (\vec{\bar I},\vec{\bar \vartheta})$ in such a
way that every next perturbation becomes the order of the square of
the preceding one: $\varepsilon V\rightarrow$ $\varepsilon^2
V_1\rightarrow$ $\varepsilon^4 V_2\rightarrow$ $\ldots
\varepsilon^{2^n} V_n$. After two successive canonical
transformations the Hamiltonian takes the form:
\begin{widetext}
\begin{multline}\label{new_H_1}
  \bar{\mathcal{H}} = \Bigl\{m_e^2c^4 + \mathcal{E}^2\beta_\|^2\Bigl[\Delta_u^2
                                       \Bigl(1+\frac{\varepsilon^2\cosh^2[k_u y_0]}{2(\Delta_u^2-\sigma_0^2)}\Bigr)
  + 2\Delta_c\sigma_0 \Bigl(1 +
              \frac{\varepsilon^2(\Delta_u^2+2\Delta_u^2\sigma_0^2-\sigma_0^2)\cosh^2[k_u y_0]}{4\sigma_0^2(\Delta_u^2-\sigma_0^2)^2}
\\
              +\frac{\varepsilon^2\Delta_u^2\sinh^2[k_u y_0]}{4\sigma_0^2(\Delta_u^2-4\sigma_0^2)}
              \Bigr) \Bigr]\Bigr\}^{1/2}
\end{multline}
\end{widetext}
Here $\Delta_u = k_u c\bar{I}_u/(\beta_\|\mathcal{E})$ and $\Delta_c
= k_u c\bar{I}_c/(\beta_\|\mathcal{E})$ are the dimensionless
integrals of motion with an accuracy $O(\varepsilon^2)$,
$\sigma_0=\omega_c/\omega_u$. The oscillation frequencies are:
\begin{equation}\label{freq_new}
\begin{split}
 \Omega_c \equiv \frac{\partial \bar{\mathcal{H}}}{\partial \bar I_c} & =
            \omega_c \Bigl[1 + \frac{\varepsilon^2\sinh^2[k_u y_0] \Delta_u^2}{4\sigma_0^2(\Delta_u^2 - 4\sigma_0^2)^2}+
\\
            &\hspace{0.9cm}\varepsilon^2\cosh^2[k_u y_0]
                               \frac{\Delta_u^4 + 2\Delta_u^2\sigma_0^2-\sigma_0^4}
                                           {4\sigma_0^2(\Delta_u^2 - \sigma_0^2)^2}\Bigl],
\\
 \Omega_u \equiv \frac{\partial \bar{\mathcal{H}}}{\partial \bar I_u} & =
          \omega_u\Delta_u \Bigr[1 - \frac{2\varepsilon^2\Delta_c\sigma_0 \sinh^2[k_u y_0]}{\Delta_u^2 - 4\sigma_0^2}-
\\
         &\hspace{0.25cm}\frac{\varepsilon^2\sigma_0^2\cosh^2[k_u y_0]}{2(\Delta_u^2  - \sigma_0^2)^2}
         \Bigl(1 + \frac{4\Delta_u^2\Delta_c}{\sigma_0(\Delta_u^2-\sigma_0^2)}\Bigr) \Bigr].
\end{split}
\end{equation}

The velocity components that are needed in the sequel are expressed
in terms of the action-angle variables as
\begin{equation}\label{traj_unperturbed}
\begin{split}
  p_x & = \frac{\varepsilon p_\|\Delta_u^2 \cosh[k_uy_0]}{\Delta_u^2-\sigma_0^2}\cos\bar\theta_u
            + p_\|\sqrt{2\sigma_0\Delta_c}\, \cos\bar\theta_c,
\\
  p_z & = \bar p_\| - \varepsilon p_\|\cosh[k_u y_0]\Bigr[\frac{\sqrt{\sigma_0\Delta_c}}{\sqrt{2}(\Delta_u-\sigma_0)}
                                                                                 \cos[\bar\theta_u-\bar\theta_c]-
\\
   &\frac{\sqrt{\sigma_0\Delta_c}}{\sqrt{2}(\Delta_u+\sigma_0)}
                                                               \cos[\bar\theta_u+\bar\theta_c]
   + \frac{\varepsilon \Delta_u\cosh[k_uy_0]}{4(\Delta_u^2-\sigma_0^2)}\cos2\bar\theta_u\Bigl].
\end{split}
\end{equation}
Here $p_\| = \mathcal{E}V_\|/c^2$ and $\bar p_\| =
p_\|\Omega_u/(k_uV_\|)$ are the initial and average axial momenta.
We have taken account of the first non-vanishing corrections only.
To complete the study we have to determine the approximate adiabatic
invariants $\Delta_u$ and $\Delta_c$. Using the relations between
old and new variables via the generating functions and initial
conditions \eqref{in_con_1} one obtains the set of equations with
respect to unknown $\Delta_u$ and $\Delta_c$. This set has the bulky
form, and we did not write it here. Instead of this it is convenient
to introduce two new auxiliary functions $\varkappa$ and $\sigma$
such that:
\begin{equation}\label{actions_1}
 \frac{\Delta_u}{\varkappa}=1+\varepsilon^2\frac{\sigma ^2\cosh ^2[k_u y_0]}{2(\varkappa ^2-\sigma ^2)^2},
  \quad \Delta_c=\frac{\varepsilon^2\varkappa^4 \cosh^2[k_u y_0]}{2 \sigma \left(\varkappa ^2-\sigma^2\right)^2},
\end{equation}
where $\varkappa$ and $\sigma$ satisfied the set of equations:
\begin{equation}\label{kappa-sigma}
\begin{split}
    \frac{\sigma}{\sigma_0}  =  &1 + \varepsilon^2\Bigl(\frac{(2\varkappa^2 - \sigma^2)\cosh^2[k_uy_0]}
                                                          {2(\varkappa^2 - \sigma^2)^2}
                             +\frac{\cosh[2k_uy_0]}{4\sigma^2}+
\\
                            & \frac{\sinh^2[k_uy_0]}{\varkappa^2 - 4\sigma^2}\Bigr),
\quad
    \varkappa = 1 - \frac{\varepsilon^2\cosh^2[k_uy_0](3\varkappa^2 + \sigma^2)}{4(\varkappa^2 - \sigma^2)^2}.
\end{split}
\end{equation}
Then the frequencies~\eqref{freq_new} are expressed in terms of
unknown constants $\varkappa$ and $\sigma$ in a simple way:
\begin{equation}
  \Omega_u = \varkappa\omega_u, \quad   \Omega_c = \sigma\omega_u.
\end{equation}
The average axial momentum and velocity prove to be equal
$\bar{p}_\|=\varkappa p_{\|}$ and $\bar v_\| = \varkappa V_\|$.
Recall that $\varepsilon = \sqrt{2}\omega_\beta/\omega_u$ and
$\sigma_0 = \omega_c/\omega_u$.

Let us find the approximate solution to equation set
\eqref{kappa-sigma}. For further analysis let us assume that
$\varkappa\sim 1$ and $\sigma\sim \sigma_0$. Then, in case of
$\sigma_0\ll 1$ and $\sigma_0\gg 1$ we may take $\varkappa=1$ and
$\sigma_0$ in the right-hand sides of Eqs.~\eqref{kappa-sigma} to
obtain the explicit solution. To consider the case $\sigma_0\sim 1$
we introduce a new small magnitude $\mu = \varkappa-\sigma$, $\mu\ll
\varkappa$. Neglecting $\mu$ in such expressions $(\mu+\varkappa)$
we get the cubic equation with respect to $\mu$:
\begin{equation}\label{cubic_eq_for_frequincies_new}
       \mu^3+c_1\mu^2+c_2\mu +c_3=0,
\end{equation}
where $c_1=\sigma_0-1$, $c_2=\varepsilon^2\sigma_0\cosh^2[k_u
y_0]/4$, $c_3=\varepsilon^2(2+\sigma_0)\cosh^2[k_u y_0]/8$. The
discriminant analysis $D(\varepsilon,\sigma_0,y_0) = p^3/27 + q^2/4$
($p=-c_1^2/3+c_2$, $q=2c_1^3/3^3-c_1c_2/3+c_3$) of cubic equation
\eqref{cubic_eq_for_frequincies_new} shows that
$D(\varepsilon,\sigma_0,y_0)<0$  in the region
$\sigma_0<\sigma_0^{crit1}$, and $D(\varepsilon,\sigma_0,y_0)>0$ in
the region $\sigma_0>\sigma_0^{crit1}$. The quantity
$\sigma_0^{crit1}(\varepsilon,y_0)$ is the solution to equation
$D(\varepsilon,\sigma_0,y_0)=0$ and equals:
\begin{multline}\label{sigma_0_crit1}
        \sigma_0^{crit1}=1 - \frac{1}{2}\Bigl(\frac{9\varepsilon\cosh[k_u y_0]}{2}\Bigr)^{2/3}-
\\
    \frac{3}{4}\Bigl(\frac{\varepsilon\cosh[k_u y_0]}{6}\Bigr)^{4/3}+\frac{7\varepsilon^2}{18}.
\end{multline}
In the region $\sigma_0<\sigma_0^{crit1}$ the solution of
equation~\eqref{cubic_eq_for_frequincies_new} has the following
form:
\begin{equation}
   \mu =  -(1-\sigma_0)/3 - 2\sqrt{-p/3}\,\cos\bigl[(\alpha+2\pi)/3\bigr],
\end{equation}
where $\cos\alpha = -q/[2\sqrt{-(p/3)^3}$]. In the region
$\sigma_0\geq \sigma_0^{crit1}$ the solution of
equation~\eqref{cubic_eq_for_frequincies_new} reads:
\begin{equation}\label{Delta_second_zone}
  \mu=  -\frac{1-\sigma_0}{3}-\frac{p}{3(\sqrt{D}-q/2)^{1/3}}+(\sqrt{D}-q/2)^{1/3}.
\end{equation}
For $\sigma_0<\varepsilon$ the above-mentioned  $\varepsilon$
expansion is not quite correct and this case should be treated
separately. The analysis shows that the trajectories remain
unchanged but to calculate the cyclotron frequency we have to make
use of another formula $\Omega_c = \{\omega_c^2 + \omega_\beta^2
\cosh^2[k_u y_0]\}^{1/2}$.
\begin{figure}[!b]
\centering %
\scalebox{1.0}{\includegraphics{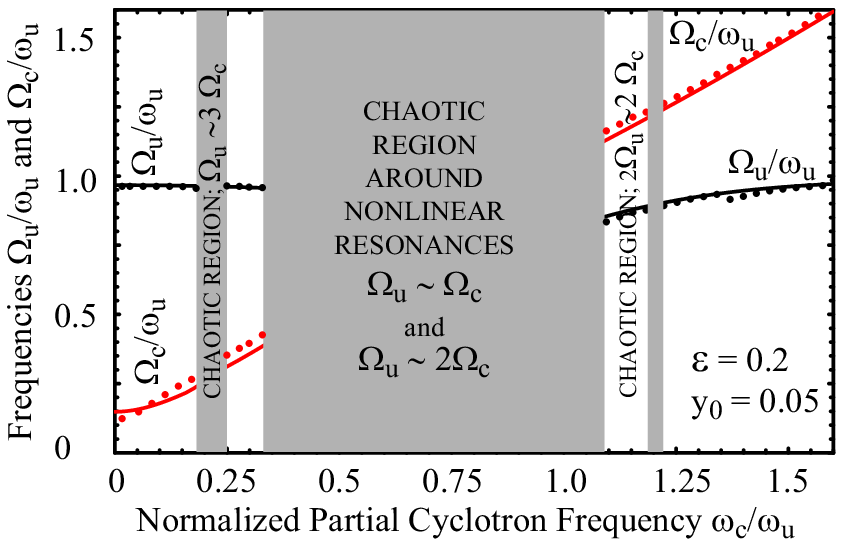}}
\caption{\label{k&s_Y_005_e_02_NoneAdiabat} Undulator and cyclotron
frequencies vs. the partial cyclotron frequency. The adiabatic
undulator entrance of electrons to the interaction region is
neglected. Solid lines are for the analytical results, while dots
correspond to the results of the numerical simulation.}
\vspace{0.5cm} \scalebox{1.0}{\includegraphics{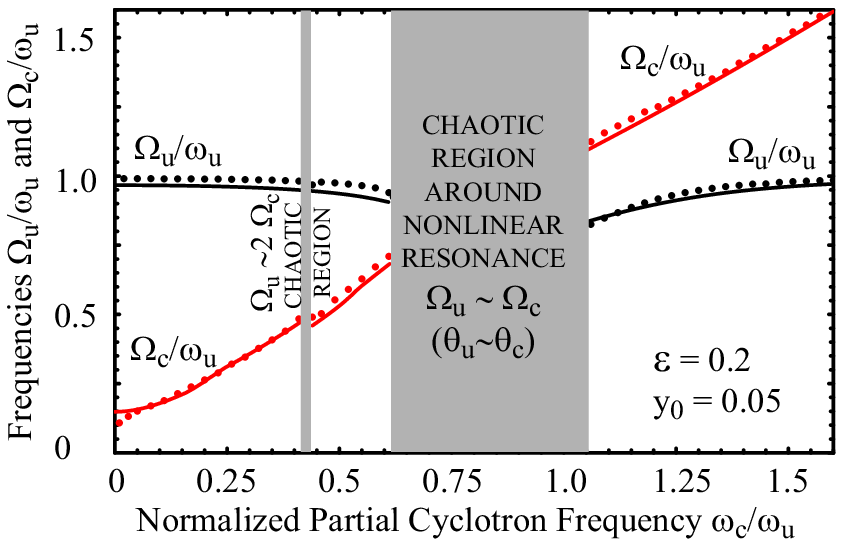}}
\caption{\label{k&s_Y_005_e_02_Adiabat} Undulator and cyclotron
frequencies vs. the partial cyclotron frequency. The adiabatic
undulator entrance is taken into account.}
\end{figure}

The comparison of the results for $\Omega_u$ and $\Omega_c$ obtained
by using the analytical expressions (solid lines) and the numerical
simulation of Eqs.~\eqref{eq_I_theta} (dots) is demonstrated in
Fig.~\ref{k&s_Y_005_e_02_NoneAdiabat} and
\ref{k&s_Y_005_e_02_Adiabat}. For both of the situations the
analytical and numerical results are seemed to be in a good
agrement. Note that in our analytical study we ignore the adiabatic
undulator entrance of electrons to the interaction region. The
comparison between Figs.~\ref{k&s_Y_005_e_02_NoneAdiabat} and
\ref{k&s_Y_005_e_02_Adiabat} gives a clear indication that the
adiabatic undulator entrance "improve integrability" of the
nonlinear dynamical system~\eqref{eq_I_theta} and reduce the width
of chaotic regions. In~\cite{We} it was found  that the initial
positive value for the $x$-component of the velocity leads to the
suppression of the chaotic region. A test electron acquires such a
positive average correction to $v_x$ passing through the region of
the adiabatic entrance. And, as a result, the chaotic dynamics of
the electron in the regular undulator region is partially
suppressed. A rough analytical estimate indicates that the average
correction is about one fourth of the amplitude of
$v_x$-oscillations in the regular undulator.

\subsection{Chaotic motion}

\begin{figure}[b!]
  \centering %
  \scalebox{1.0}{\includegraphics{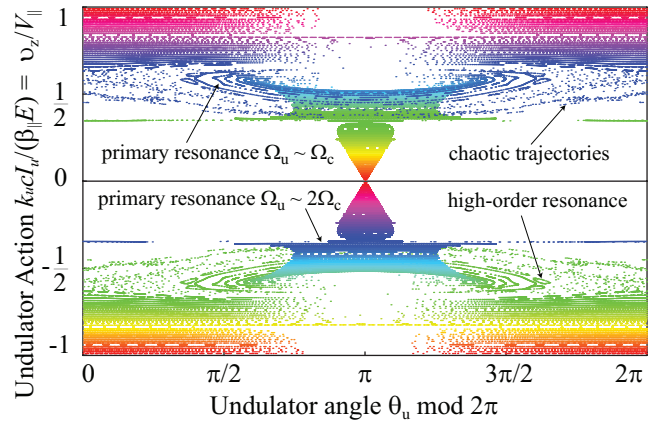}}
\caption{\label{Poincare} Poincare mapping to the
system~\eqref{eq_I_theta}; $\vartheta_c=0$.}
\end{figure}
The equation set \eqref{eq_I_theta} has a lot of nonlinear
resonances~\eqref{resonances} ($n \vartheta_u\approx m \vartheta_c$)
between the undulator and cyclotron degrees of freedom, therefore
one can expect the appearance of chaotic dynamics in the system
behavior. In Fig.~\ref{Poincare} we have demonstrated the Poincare
mapping in which the primary and the higher resonances are seen; the
separate dots correspond to the stochastic trajectories. Note that
the average axial velocity for the majority of the stochastic
trajectories equals zero. As evident from Eq.~\eqref{actions_1} and
Eq.~\eqref{kappa-sigma}, the undulator action (integral of motion!)
can vanish under some conditions. This means the destruction of the
integral motion~\cite[ch.~5]{Zaslavskii_1988} and chaotization of
the dynamics of the test electron. Hence, the motion becomes
stochastic if the difference between the undulator and cyclotron
frequencies becomes less than the betatron frequency
\begin{equation}\label{chaos_condit}
  |\Omega_u-\Omega_c| \leq \sqrt{2}\omega_\beta\cosh[k_u y_0]
   \quad (|\kappa-\sigma|\leq \varepsilon \cosh[k_u y_0]).
\end{equation}
Such a criteria was initially proposed in~\cite{We} using some
numerical findings. With the derived $\varkappa$ and $\sigma$ we get
the expression describing the location of the region of the dynamic
chaos:
\begin{equation}\label{boundary_chaos}
\begin{split}
  &\sigma_0^{crit1}\leq \sigma_0 \leq\sigma_0^{crit2},
\\
 &\sigma_0^{crit2}=\frac{2}{3}+\frac{28\varepsilon \cosh[k_u y_0]}{27}
                                                     + \frac{5\varepsilon^2\cosh^2[k_u y_0]}{18}.
\end{split}
\end{equation}
It turns out that the chaotization condition is inconsistent with
the solution of equation set \eqref{kappa-sigma} for $\varepsilon$
less than the minimal value of $\varepsilon^{min}$
\begin{equation}
  \varepsilon^{min}\cosh[k_u y_0] = 0.0786.
\end{equation}
In the used approximations this implies that there is no the chaotic
region for $\varepsilon<\varepsilon^{min}$.
\begin{figure}[t!]
  \centering %
  \scalebox{1.0}{\includegraphics[230,103][468,336]{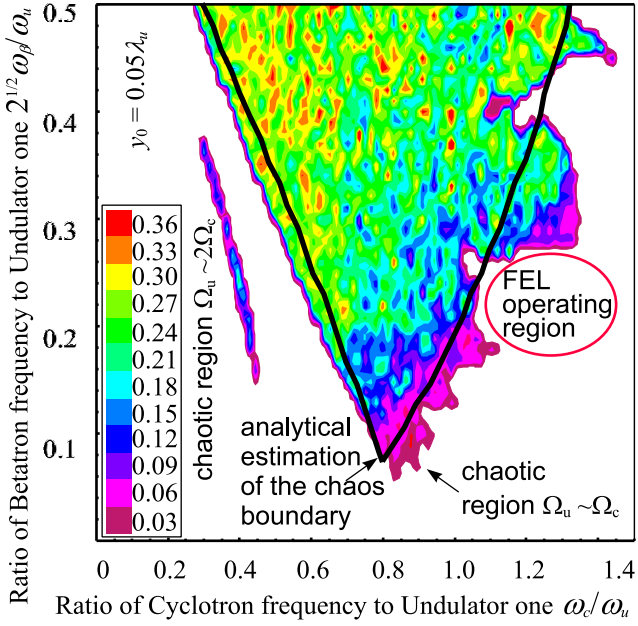}}
\caption{\label{Lyap_map}  The major Lyapunov exponent map. Black
solid lines show the boundaries of the chaotic region according to
the analytical formulae~\eqref{boundary_chaos}.}
\end{figure}
In Fig.~\ref{Lyap_map} we have illustrated the results of the
numerical calculations for the major Lyapunov exponent. The solid
lines calculated using equation~\eqref{boundary_chaos} show the
boundaries of the chaotic region. They are in a good agreement with
the results of numerical simulation.

In what follows we analyze the studied FEL within the region of the
{\it{regular}} dynamic.

\section{The reduced model of the FEL}

In this section we construct the time-dependent reduced FEL model
that allows for the complicated dynamics of electrons in the pump
magnetic field and intramode scattering in an irregular waveguide.
Since we are going to analyze the resonant interaction between the
microwave and electron beam we may approximately rewrite
vector-potential $A_x^s$ as (see~\ref{appendix2})
\begin{multline}\label{microwave_split}
  A_x^s = \mathrm{Re}\Bigl\{ \bigl[V_+(z,t) + V_-(z,t)\bigr]\sqrt{\frac{k_z^0b}{k_z(z)w(z)}}\times
\\
       \cos\Bigl(\frac{\pi y}{w(z)}\Bigr) \,e^{-i\omega t+i\Psi(z)}\Bigr\}.
\end{multline}
Here $\Psi(z) = \int_0^z k_z(z')dz'$, $V_+(z,t)$ and $V_-(z,t)$ are
the amplitudes of the forward and backward waves governed by the
following excitation equations
\begin{subequations}\label{time_amplitude_split}
\begin{align}
\label{time_amplitude_split1}
    &\frac{\partial V_+}{\partial z} + \frac{1}{v_{gr}(z)}\frac{\partial V_+}{\partial t} =
             J^{e}(z,t),
  \\
\label{time_amplitude_split2}
   &\frac{\partial V_-}{\partial z} - \frac{1}{v_{gr}(z)}\frac{\partial V_-}{\partial t} + 2i k_z V_- =
                                          \frac{\partial k_z}{\partial z}\frac{V_+}{2k_z}.
\end{align}
\end{subequations}
The Eq.~\eqref{time_amplitude_split1} describes the interaction
between the beam and forward wave, whereas the
Eq.~\eqref{time_amplitude_split2} describes the scattering of the
forward wave to the backward one. The boundary conditions are
\begin{equation}
  V_+\bigr|_{z=0} = V_0, \quad V_-\bigr|_{z=L} = 0.
\end{equation}
The effective current
\begin{multline}
  J^{e}(z,t)= \frac{4i I_0\omega}{k_z^0Sc} \sqrt{\frac{k_z^0 b}{k_z w}}
                       \frac{1}{Y_b} \int\limits_{-Y_b/2}^{Y_b/2}\!\! dy_0
                        \int\limits_{t - t^p-\pi/\omega}^{t - t^p+\pi/\omega}\!\! dt_e
\times
\\
   \frac{p_x (z;y_0,t_e)}{p_z(z;y_0,t_e)}
  \cos\Bigl(\frac{\pi y(z;y_0,t_e)}{w(z)}\Bigr) e^{i\omega t(z;y_0,t_e)-i\Psi(z)}
\end{multline}
at moment $t$ and position $z$ is generated by the group of
electrons entered into the interaction region during the time
interval from $(t - t^p(z;y_0)-\pi/\omega)$ to $(t -
t^p(z;y_0)+\pi/\omega)$. Here $t^p(z;y_0)$ is the arrival time of
the electron, which moves in the pump magnetic
field~\eqref{pump_field}, to the cross-section $z$. According to
Eq.~\eqref{nonregular_amplitude_time} and Eq.~\eqref{charge_density}
the integrating over $t'$ yields a non-zero result only if $t -
\pi/\omega\leq t(z;y_0,t_e) \leq t + \pi/\omega$ because of the
Dirac delta. Since the right-hand side of excitation
equation~\eqref{nonregular_amplitude_time} is a slow function of
time we may write $t(z;y_0,t_e)\approx t_e + t^p(z;y_0)$ and find
integration limits with respect to $t_e$.

In the previous Section we have studied the nonlinear system with
the Hamiltonian~\eqref{Hamiltonian_upperturbed} and found
trajectories as functions of actions $\vec{\bar I}$ and angles
$\vec{\bar\vartheta}$. Now let us take into account the
ponderomotive potential $W(\vec r,\vec P,t)$. We will hold
$\bar{\mathcal{H}}(\vec r, \vec P)$ as the integrable part of the
Hamiltonian~\eqref{Hamiltonian_perturbed} and consider the
relation~\eqref{traj_unperturbed} between $(\vec r, \vec P)$ and
$(\vec{\bar I},\vec{\bar\vartheta})$ as a variable replacement rule,
regarding $(\vec{\bar I},\vec{\bar\vartheta})$ as new unknown
variables. The perturbation $W(\vec{\bar I},\vec{\bar\vartheta},t)$
periodically depends on angles $\bar\vartheta_u$ and
$\bar\vartheta_c$, therefore it can be represented as a double
Fourier-series over $\bar\vartheta_u$ and $\bar\vartheta_c$:
$W\propto W_{n,m} e^{-i\omega t + i\Psi(\bar\vartheta_u) +
in\bar\vartheta_u+im\bar\vartheta_u}$ ($m$ and $n$ are integers). As
we have already known from Sec.~\ref{section:Dynamics} the slower is
phase changing $\omega t -\Psi- n\bar\vartheta_u-m\bar\vartheta_u$,
the stronger is the action of Fourier-component $W_{n,m}$ (see text
below Eq.~\eqref{eq_I_theta}). The main principle of the nonlinear
resonance analysis is simple~\cite[ch.~3]{Zaslavskii_1988}: the
`troublesome' resonant term is separately extracted from the
perturbation expansion and, in the sequel, the dynamics caused by
this term is studied. Further on, we analyze the undulator resonance
\begin{equation}\label{res_con_2}
  \bar\vartheta_u + \int_0^{\bar\vartheta_u/k_u} k_z(z')dz' \approx \omega t \quad ((k_z+k_u)\bar v_\|\approx \omega).
\end{equation}
and may write the ponderomotive perturbation as
\begin{equation*}
  W(\vec{\bar I},\vec{\bar\vartheta},t) \approx \mathrm{Re}\bigl\{\tilde W(\bar I_u,z(\bar\vartheta_u),t)
                    e^{i\psi(\bar\vartheta_u,t)}\bigr\},
\end{equation*}
where
\begin{equation*}
  \tilde W = -ec\, \frac{\varepsilon p_\|\Delta_u^2 \cosh[k_uy_0]}{\Delta_u^2-\sigma_0^2}
              \bigl(V_+ + V_-\bigr)\sqrt{\frac{k_z^0b}{k_zw}}\cos\Bigl(\frac{\pi y}{w}\Bigr).
\end{equation*}
Quantity $\tilde W(\bar I_u,z(\bar\vartheta_u),t)$ is a slow
function of $\bar\vartheta_u$ and $t$ via a slow variation of the
waveguide profile and the microwave amplitude. The equations of
motion is
\begin{equation}\label{eq_I_theta_short-cut}
\begin{split}
  \dot{\bar{I}}_u & = -\frac{1}{2\mathcal{H}}\frac{\partial W}{\partial \bar\vartheta_u}
  \quad
  \dot{\bar{\vartheta}}_u = \frac{\mathcal{E}}{\mathcal{H}}\;\Omega_u(\vec{\bar I})
                           + \frac{1}{2\mathcal{H}}\frac{\partial W}{\partial \bar I_u},
\\
  \dot{\bar{I}}_c &= 0, \qquad \quad\dot{\bar{\vartheta}}_c = \Omega_c(\vec{\bar I})\;\mathcal{E}/\mathcal{H}.
\end{split}
\end{equation}
To obtain the equations of motion in the simplest form and clearly
demonstrate the physics of FELs with the axial magnetic field we
additionally suppose that $\mathcal{H}-\mathcal{E}\ll \mathcal{E}$
(`Compton limit'). Applying the method of the nonlinear
resonance~\cite[ch.~3]{Zaslavskii_1988} to
Eqs.~\eqref{eq_I_theta_short-cut}, using $z$ as a new independent
variable and defining the ponderomotive phase as $\psi =
 k_u z + \int_0^z k_z(z')dz' - \omega t$, we get the equation for $\psi$
\begin{equation*}
 \frac{\partial^2 \psi}{\partial z^2} -\frac{\partial k_z}{\partial z}= -\frac{1}{\bar v_\|^2}\frac{\omega}{\Omega_u}
            \Bigl(1+\frac{k_z}{k_u}\Bigr)\frac{1}{2\mathcal{H}} \frac{\partial W}{\partial \psi}
            \frac{\partial}{\partial {\bar{I}_u}} \Bigl[\frac{\mathcal{E}}{\mathcal{H}}\;\Omega_u\Bigr],
\end{equation*}
where the derivative of the undulator frequency is
\begin{equation*}
  \frac{\partial}{\partial {\bar{I}_u}} \Bigl[\frac{\mathcal{E}}{\mathcal{H}}\;\Omega_u \Bigr] \approx
              \frac{\omega_u\mathcal{E}}{I_u^0}
              \frac{\partial}{\partial {\bar{I}_u}}\frac{{\bar{I}_u}}{\mathcal{H}}
                \approx
              \frac{\omega_u^2}{\beta_\|^2 \bar\gamma_\|^2 \mathcal{E}}
    \quad \Bigl(\frac{1}{\bar\gamma_\|^2} = 1-\frac{\bar v_\|^2}{c^2}\Bigr).
\end{equation*}
Expanding the ponderomotive current into a series with respect to
angles and taking into account only the resonant term in the
excitation equation one can write the reduced FEL model in the
following manner:
\begin{equation}\label{reduced_model}
\begin{split}
  &\frac{\partial^2\tilde\psi}{\partial \zeta^2} = -Y(y_0)Z(\zeta)
                                 \;\mathrm{Re}\{\bigl(F_+ + F_-\bigr)  e^{i\tilde\psi} \},
\hspace{-1cm}
  \\
    &\frac{\partial F_+}{\partial \zeta}-i\delta_z(\zeta)F_+
           + \Bigl(\frac{\bar v_\|}{v_{gr}}-1\Bigr)\frac{\partial F_+}{\partial \tau}
= \tilde J^{e}(\zeta,\tau),
  \\
   &\frac{\partial F_-}{\partial \zeta}+i(2k_z\ell_g-\delta_z)F_-
             - \Bigl(1+\frac{\bar v_\|}{v_{gr}}\Bigr)\frac{\partial F_-}{\partial \tau}
                  = \frac{\partial k_z}{\partial \zeta}\frac{F_+}{2k_z},\hspace{-1cm}
\\
      & \tilde J^{e}(\zeta,\tau) =  \frac{Z(\zeta)}{\pi Y_b}\!\! \int\limits_{-Y_b/2}^{Y_b/2}\!\!
                           dy'_0                \!\!
                        \int\limits_{-\pi}^{\pi}\!\!     d\psi'_0\, Y(y'_0) e^{-i\tilde\psi(\zeta,\tau;y'_0,\psi'_0)},
\hspace{-1cm}
\\
  &\tilde\psi|_{\zeta=0} = \psi_0, \quad \frac{\partial \tilde\psi}{\partial\zeta}\Bigr|_{\zeta=0}
                  \equiv     \delta_y(y_0) =
                   \frac{\ell_g\omega}{\bar v_\|(0)}-\frac{\ell_g\omega}{\bar v_\|(y_0)},
\\
  & F_+\Bigr|_{\zeta=0} = F_0, \quad F_-\Bigr|_{\zeta=L/\ell_g} = 0.
\end{split}
\end{equation}
Here  $\zeta = z/\ell_g$ and $\tau=(\bar v_\| t-z)/\ell_g$ are the
dimensionless longitudinal coordinate and "retarded time"~
\cite{Bonifacio_1992};
\begin{equation}
  F_\pm = - \frac{iec\omega^2 \ell_g^2}{\sqrt{2}\bar v_\|^3\bar\gamma_\|^2\mathcal{E}}
       \frac{\omega_\beta}{\Omega_u} \, V_\pm \,e^{i\int_0^\zeta\delta_z(\zeta')d\zeta'}
\end{equation}
is the normalized field amplitude;
$\tilde\psi(\zeta,\tau;y_0,\psi_0) = \psi(\zeta,\tau;y_0,\psi_0) -
\int_0^\zeta\delta_z(\zeta')d\zeta'$ is the ponderomotive phase;
$\psi_0$ and $y_0$ are the initial entrance phase and the initial
transverse displacement of the electron's position from the
undulator symmetry plane $y=0$;
\begin{equation}
  \ell_g^{-3} = \frac{1}{\bar \beta_\|^3 \bar\gamma_\|^3}\frac{I_0}{I_\alpha}
          \frac{2\pi \omega^2}{k_z^0Sc^2}\frac{\omega_\beta^2}{\Omega_u^2},
\end{equation}
where $I_\alpha = m_ec^3/e\approx -17$~kA is the Alfv\'{e}n current
(recall that $e,I_0<0$). The parameter $ \ell_g$ is called the gain
length~\cite{Bonifacio_1992} (the spatial growth rate of the FEL
without the axial magnetic field is equal to $\ell_g^{-1}$ for zero
detuning). The explicit dependence of the reduced FEL model on the
transverse electron's position and the axial position are given by
the relations
\begin{equation*}
  Y = \frac{\Omega_u^2(y_0)\cosh[k_uy_0]}{\Omega_u^2(y_0)-\Omega_c^2(y_0)}\cos\Bigl[\frac{\pi y_0}{w(z)}\Bigr],
  \;
   Z=\sqrt{\frac{k_z^0 b}{k_z(z) w(z)}}.
\end{equation*}
The dimensionless detuning parameter is
\begin{equation}
  \delta_z(\zeta) = \ell_g\Bigl(k_z(\zeta)+k_u-\frac{\omega}{{\bar v_\|}(0)}\Bigr).
\end{equation}
Note that our model includes two detuning parameters:
$\delta_z(\zeta)$ changing along the interaction region  and
$\delta_y(y_0)$ changing across the beam. The physical meaning of
these parameters is discussed in detail below.

The time-dependent model~\eqref{reduced_model} allows for the
intricate dynamics of electrons in the pump magnetic
field~\eqref{pump_field}, the effect of the electron beam finite
thickness and the intramode scattering in the profiled waveguide
(the intramode scattering  acts actually as a feedback). Our
model~\eqref{reduced_model} is exactly coincident with the
Colson-Bohifacio model~\cite{Colson_1976,Bonifacio_1984} if a
free-space case is employed, the axial magnetic field equals zero,
the electron beam is ideally thin and ultrarelativistic. In the
above case, the field amplitude $F_+$ depends mainly on detuning
parameter $\delta_z$ if the initial value of $F_+$ is sufficiently
small. Then the FEL operates efficiently if $|\delta_z|<2$ (see
Fig.~2 in~\cite{Bonifacio_1992,Ninno_2004}). Otherwise, our
model~\eqref{reduced_model} is dependent upon more than one
parameter and incorporates some novel effects. It is worth noting
that using the method of nonlinear resonance in the action-angle
phase space one can reduce any Hamiltonian with the resonant
perturbation to the so-called Universal Hamiltonian of nonlinear
resonance~\cite[ch.~3]{Zaslavskii_1988}. This means that any
resonant beam-wave interaction can be described within the framework
of the reduced model that includes the pendulum-like equations of
motion of electrons and the excitation equation of Colson-Bonifacio
type. The equations of motion have dimensions of one and a half. In
our case, the beam-wave energy transfer occurs through the undulator
degree of freedom only, whereas the energy stored in the cyclotron
degree remains unchanged.

What is important is that the model~\eqref{reduced_model} depends
solely on the average axial velocity via the detuning but does not
depend on the particular scalar components of the initial velocity.
This results in that the FEL efficiency is only dictated by the
spread of the average axial velocity such that $\delta\bar v_\|
\propto \delta v_z + \varepsilon \delta v_x + \varepsilon^2\delta
v_y$, where $\delta v_i(t_e)$ $(i=x,y,z)$ is the magnitude of the
initial velocity spread ($\delta v_i/\bar v_\|\ll 1$) and
$\delta\bar v_\|(t_e)$ is the average axial velocity spread. Recall
that $\varepsilon = \sqrt{2}\omega_\beta/\omega_u$ is a small
parameter. It is clear that one first needs to minimize the initial
axial velocity spread. The analysis of the FEL indicates that the
velocity spread changes the efficiency insignificantly if the
detuning caused by the spread is much smaller than
unity~\cite{Bonifacio_1992}. For the ideally thin beam ($Y_b
\rightarrow 0$) it yields the condition
\begin{equation}
  \frac{(k_z^0+k_u)\ell_g}{Y^{2/3}} \frac{\mu_{\bar v_\|}}{\bar v_\|} \ll   1,
\end{equation}
where $\mu_{\bar v_\|}$ is the variance of $\delta\bar v_\|$.  It
was shown in~\cite{Freund_1986} (see also the results of the
numerical simulation in~\cite{We}) that essential decreases in the
sensitivity of the the efficiency to the initial beam spread can be
obtained if the undulator frequency is close to the cyclotron one
(multiplier $Y^{-2/3}$ attains its minimal value).

In general, the ponderomotive potential enhances as the undulator
frequency tends to the cyclotron one ($Y(y_0)$ increases). This
results in a stronger coupling between the wave and electrons. Such
an effect referred to as the magnetoresonance is well-known in the
literature~\cite{Sprangle_1978} and the recent detailed
study~\cite{We} confirmed the usefulness of such a regime for a
planar FEL configuration. However, the magnetoresonance effect is
not so effective when the beam has a finite thickness. Electrons
with the different initial transverse positions, $y_0$, undergo the
action of the different magnitudes of the pump magnetic
field~\eqref{pump_field}. Then the average velocity of the electron
depends on its initial transverse positions $y_0$ (this dependence
particulary strong near the magnetoresonance). At the same time the
average velocity of the electron of the beam governs the initial
`transverse' detuning $\delta_y(y_0)$ between the electron and the
wave. Hence, the value of $\delta_y(y_0)$ changes across the beam,
and the contribution of different electrons to the total efficiency
might be quite different. To demonstrate this effect we simulate
Eqs.~\eqref{reduced_model} for the parameters close to the
experiment~\cite{Destler_1996} and assume there is additional axial
magnetic field 20~kG as well.
\begin{figure}[tb!]
  \centering %
  \scalebox{1.0}{\includegraphics[665,470][903,600]{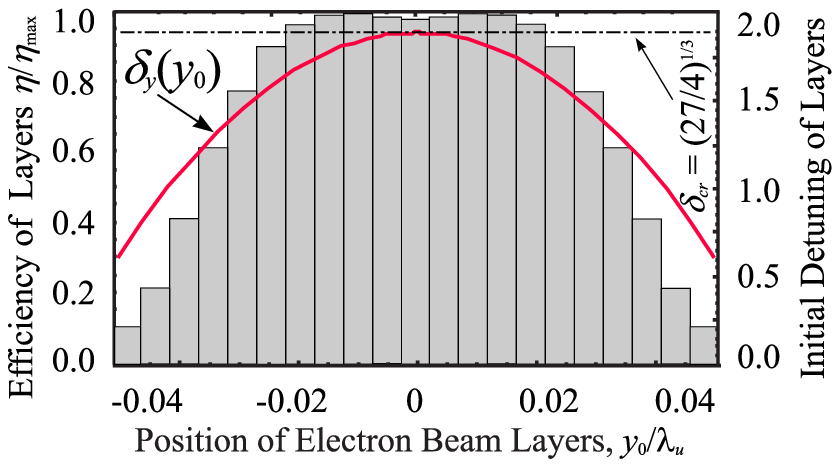}}
\caption{\label{FEL_y0} The relative efficiency (histogram) and the
initial detuning (red solid line) of electron beam layers vs. the
normalized transverse displacement of layers from the symmetry plane
$y=0$.}
\end{figure}
In the simulation we split the beam into 21 layers in the transverse
cross-section. Each layer is also uniformly distributed into 50
macroparticles entering within one wave period. Recall that the
physical system under study is homogenous in the $x$-direction. The
results are shown in Fig.~\ref{FEL_y0}. The internal layer operates
in the regime of optimal (with respect to efficiency) detuning
$\delta_z(\zeta) =\delta_{cr} = (27/4)^{1/3}$ and $\delta_y\equiv 0$
but the external layers operate with the non-optimal detuning
because of the variation in $\delta_y$ (illustrated in
Fig.~\ref{FEL_y0} as the solid line) across the beam.

It should be anew mentioned that under certain conditions the
integral of motion of a typical electron fails and dynamics becomes
chaotic. Now we can derive the chaotization condition including the
effect of the microwave. In the microwave saturation region one can
hold the average value of the undulator action as an integral of
motion, then the condition $\langle I_u \rangle>0$ has to be
fulfiled to preserve the regular dynamics of electrons and the
validity of model~\eqref{reduced_model}. In the case of the
steady-state regime, when the beam is thin and the waveguide is
regular, the improved chaotization condition is
\begin{equation}\label{chaos_condit_mod}
   \frac{2\omega_\beta^2}{(\Omega_u-\Omega_c)^2} + \frac{\bar\gamma_\|^2\Omega_u}{\omega}
                 \frac{|F_+|^2}{8k_u\ell_g}\Bigl|\frac{\Omega_u}{\Omega_u-\Omega_c}\Bigr|^{2/3}<1.
\end{equation}
Here we have considered that the microwave field modifies the
undulator action by the additional quantity
\begin{equation*}
  \Delta I_u = \frac{\dot{\tilde\psi} \Omega_u}{\omega} \Bigl(\frac{\partial \Omega_u}{\partial \bar I_u}\Bigr)^{-1}.
\end{equation*}
Besides, we used the constant of the motion to the Colson-Bonifacio
model~\cite{Krinsky_1993}
\begin{equation*}
  \langle\frac{d\tilde\psi}{d\zeta}\rangle = -\frac{|F_+|^2 - |F_0|^2}{4}.
\end{equation*}
It is obvious from Eq.~\eqref{chaos_condit_mod} that the microwave
may cause the chaotization of electrons even if the dynamics was
regular in the pure pump field.

Another important feature of the model~\eqref{reduced_model} is that
it takes into account the effect of waveguide profiling. This effect
exhibits the coupling between the forward and backward waves because
of intramode scattering as well as the dependence of the wave
number, $k_z$, on the axial position $z$ through the varying
waveguide width, $w(z)$. As a result, the detuning $\delta_z$ is
also a function of the axial position and its control can be used to
govern the beam-wave interaction.

\section{Control of the beam-wave interaction: FEL with the \protect\\ optimized waveguide profile}

Now discuss the physical principle of the control of the beam-wave
interaction.  For simplicity we consider the steady-state regime and
the thin beam. We also neglect the backward wave generation assuming
that $w(z)$ is a slow function of $z$. Rewriting the complex
amplitudes of the wave and ponderomotive current as $F_+ =
|F_+|e^{i\alpha}$ and $\tilde J^{e} = |\tilde J^{e}|e^{iu_\ast}$ and
using Eqs.~\eqref{reduced_model} we arrive at the system:
\begin{equation}\label{system_Soltnzev}
\begin{split}
  &  \frac{d^2\tilde\psi}{d\zeta^2} = - |F_+|\cos(\alpha+\tilde\psi),
     \quad \frac{d|F_+|}{d\zeta}= |\tilde J^{e}|\cos(\alpha - u_\ast),
    \hspace{-1cm}
\\
  & \frac{d\alpha}{d\zeta} = \delta_z \bigr(\zeta) -|\tilde J^{e}|\sin(\alpha - u_\ast)/|F_+|.
  \\
  &  u_\ast = \mathrm{Arg}\Bigl(\frac{1}{\pi}\int_0^{2\pi}\!\!\! e^{-i\tilde\psi}\,d\psi_0\Bigr),
     \quad |\tilde J^{e}| = \frac{1}{\pi}\Bigl|\int_0^{2\pi}\!\!\! e^{-i\tilde\psi}\,d\psi_0\Bigr|.
     \hspace{-1cm}
\end{split}
\end{equation}
\begin{figure}[tb!]
  \centering %
  \scalebox{1.0}{\includegraphics[236,121][462,318]{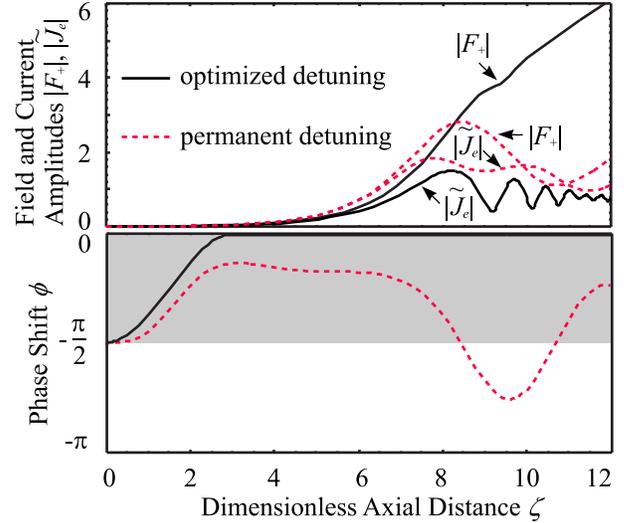}}
\caption{\label{optimal_phase_shift} The simulation results for the FEL with and without the phase shift optimization are demonstrated.}
\end{figure}
Here $\tilde\psi$, $|F_+|$ and $\alpha$ are the unknown quantities
governed by the differential equations, and $u_\ast$ and $|\tilde
J^{e}|$ are given by definition. The phase of the current $u_\ast$
defines the position of the bunch center in the system of
coordinates moving with the velocity of the
beam~\cite[p.~160]{VanSoln_1973} (see
also~\cite[p.~325]{Tsimring_2007}). One can see that if the phase
shift between the current and the wave, $\phi=\alpha - u_\ast$,
belongs to the interval from $-\pi/2$ to $\pi/2$, then the
right-hand side of the equation governing wave amplitude $|F_+|$
(second equation in the upper line of~\eqref{system_Soltnzev}) has a
positive sign and the amplitude itself grows. The phase shift $\phi$
governs the energy transfer from the beam to the wave (or v.v.)
because the local interaction power is $dP/d\zeta \propto |\tilde
J^{e}||F_+|\cos\phi$. This implies that we can increase the
efficiency by controlling $\phi$ along the interaction region by
changing the detuning parameter $\delta_z(\zeta)$ in an appropriate
way. The idea of such an optimization was originally proposed in the
TWT theory~\cite{Solnzev_1971}. Here, for example, we demonstrate
the simple indirect optimization method~\cite{Solnzev_1971}. Now
assume that the phase shift, $\phi$, satisfies the relation:
\begin{equation}\label{phase_difference}
  \begin{split}
     \alpha - u_\ast \equiv \phi_{opt}(\zeta)   &= - \pi \sin^2\frac{\pi \zeta}{2 L_{pr}}, \quad \zeta\leq L_{pr},
    \\
     \alpha - u_\ast \equiv \phi_{opt}(\zeta)  &=  0, \qquad\qquad\qquad \zeta > L_{pr},
  \end{split}
\end{equation}
where $L_{pr}$ is the start point of the region with the permanent
value of $\phi$. Then we have to find $\tilde\psi$ and $|F_+|$ using
Eqs.~\eqref{system_Soltnzev} and Eq.~\eqref{phase_difference} (in
the right-hand sides of Eqs.~\eqref{system_Soltnzev} the expression
$(\alpha - u_\ast)$ should be replaced by $\phi_{opt}$). Now we can
restore the information about the waveguide profile using the
equation for the detuning parameter that follows
from~\eqref{system_Soltnzev}:
\begin{equation}
  \delta_z \bigr(\zeta) = \frac{d u_\ast}{d\zeta} + \frac{d \phi_{opt}}{d\zeta} + (|\tilde J^{e}|\sin \phi_{opt})/|F_+|.
\end{equation}
The results from the calculation of the amplitudes of the wave
$|F_+|$ and the current $|\tilde J^{e}|$, and the phase shift $\phi$
are shown in Fig.~\ref{optimal_phase_shift}. In this figure we also
plotted the FEL characteristics for the constant detuning. We can
see that the wave amplitude can be significantly enhanced
(efficiency increased several times). However, the demonstrated
optimization technique is useful only for a slightly improved
efficiency because  the waveguide profiles are to be rather
complicated from the practical point of view in an effort to
considerably increase the efficiency. Then more elaborated
mathematical approaches, which simultaneously allow one to control
the practical realizability of optimal waveguide profiles, should be
used. In this paper we apply some type of a genetic
algorithm~\cite{Mitchell_1999} to perform the FEL optimization. The
principle of evolutionary optimization is rather simple: we generate
a lot of waveguide profiles and then perform numerical simulation of
the reduced model~\eqref{reduced_model} using these profiles. Then
we choose the best profiles, cross and modify them, and perform the
simulation again. As a result one can find a few best profiles.
Finally we must check that these found optimal profiles are really
useful. We have to simulate the non-simplified original model
(formulated in Sec.~\ref{section:model}) using these profiles. Let
us consider the results of the optimization for a practical example.

\begin{widetext}
  \begin{figure}[tb!]
    \centering %
    \scalebox{1.0}{\includegraphics[68,88][574,351]{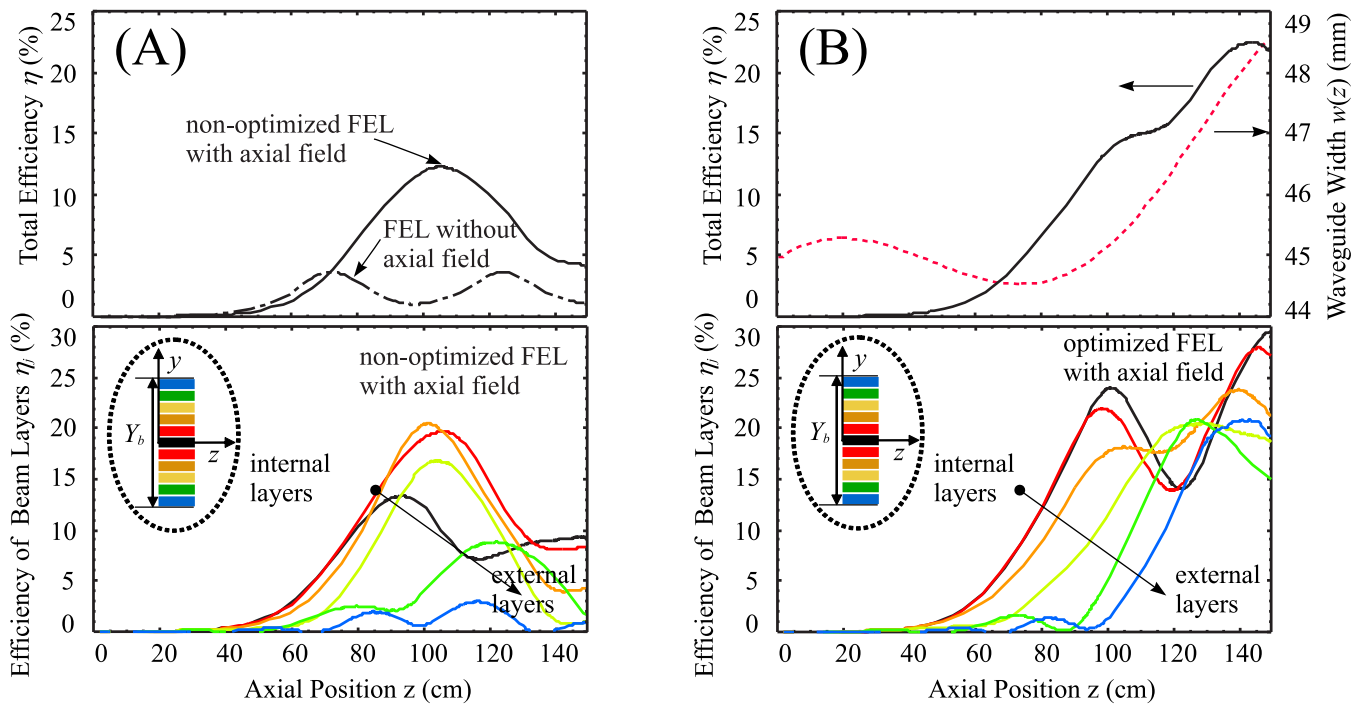}}
    \caption{\label{optimized_FEL} The FEL efficiency and waveguide width vs. the interaction length.
     The results of the non-simplified model simulation are demonstrated.}
  \end{figure}
\end{widetext}

The FEL parameters are chosen to be close to the parameters of
experiment~\cite{Destler_1996}: 450-kV beam voltage, $|I_0|=16$-A
beam current, 1.0~mm $\times$ 2.0~cm sheet electron beam interacts
with the TE$_{01}$ mode (the field varying along the narrow wall) of
the 4.5~mm $\times$ 4.0~cm rectangular waveguide. The undulator
magnitude increases adiabatically within six periods and the
undulator is characterized by parameters $B_u = 3.5$~kG and
$\lambda_u = 1.0$~cm in the regular region. A 1-kW input signal with
the 4.0~mm wavelength  is injected. In our simulation we assume that
there is the axial magnetic field 20~kG as well. The wavelength is
slightly different from that in the experiment because of the
different average axial velocity. In Fig.~\ref{optimized_FEL}A the
results for the FEL with the axial field but without optimization
are shown. Using the magnetoresonance effect we can significantly
enhance the efficiency. It was 4\% efficiency without the axial field in the
experiment and it is 12\% efficiency with the axial magnetic field. However,
there is a weak interaction between the
external beam layers and the microwave because
different layers of the electron beam have different "transverse"
detuning with the wave due to the transverse inhomogeneity of
the pump magnetic field. Geometric positions of different beam
layers at the beginning of the interaction region are shown inside
the dotted ellipse. The black curve is for the central layer. Other layers
are displaced with respect to the symmetry plane $y=0$.

In Fig.~\ref{optimized_FEL}B the results for the optimized FEL with
the axial field are presented. Using the waveguide with the optimized
profile one can double the efficiency so that the final efficiency
is around 22\%. We also see the external layers
interact with the wave much more effectively in the optimized FEL. So, by changing the waveguide
profile we control beam-wave interaction thus
increasing the FEL efficiency.

\section{Summary and Discussion}

The operation of the planar FEL-amplifier with the axial magnetic
field and the irregular waveguide is studied. The self-consistent
model, which includes the excitation equation and the equations of
motion along with the expressions for the radiated field and the
microscopic current density, is formulated. In order to find the
parameters and the waveguide profile that provides the maximal
efficiency one has to perform some optimization of the FEL. However,
the conventional numerical optimization methods fail to work because
a vast amount of computational resources is required. Typically,
about several thousand equations of motions and the partial
differential equation for the wave amplitude have to be simulated.
In this paper I propose another approach to the problem. The
investigation is divided into several stages: initially I partly
integrate equations of motion and the excitation equation in an
analytical way using methods of nonlinear dynamics. As a result, the
universal reduced FEL model is derived in special phase space. Then
with this model and some principles of evolutionary computations
(genetic algorithms) I perform the numerical optimization of the
waveguide profile. Finally, the simulation of the non-simplified
original model using the found optimal waveguide profiles is carried
out. So, one can come closer to understanding of what increase in
the efficiency can be achieved in practise.

\begin{figure}[tb!]
    \centering %
    \scalebox{1.0}{\includegraphics[269,150][429,289]{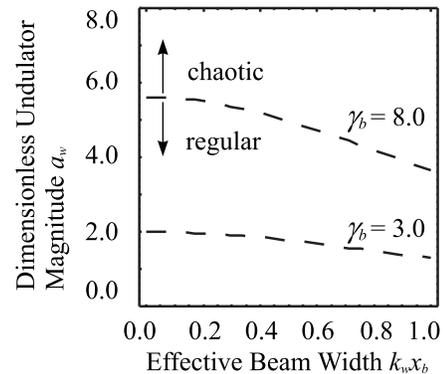}}
    \caption{\label{fig_chaos_undul} The boundaries of regular and chaotic regions
                               obtained using Eq.~\eqref{chaos_pure_undul}
                              (cf. with Fig.~2 in~\cite{ChenDavidson1990}).}
\end{figure}

To derive the reduce model one first can find the integrals of
motion of a test electron in the pump magnetic
field~\eqref{pump_field} applying the overconvergent method. The
dynamics is completely governed by the two integrals of motion
corresponding to the undulator and cyclotron degrees of freedom. At
the same time it is reasonable to use other two parameters that
completely define the dynamics as well: the first governing
parameter $\varepsilon=\sqrt{2}\omega_\beta \cosh[k_u y_0]/\omega_u$
is the level of nonlinearity and the second one
$\sigma_0=\omega_c/\omega_u$ shows how the system is close to the
magnetoresonance. The complete description of the dynamics is given
in terms of these parameters. It is well known that the dynamics in
the pump field is chaotic for some parameters~\cite{Buzzi_1993}, so
the explicit expression~\eqref{boundary_chaos} describing the region
location of the dynamic chaos in the parameter space ($\varepsilon,
\sigma_0$) are derived and the existence domain of integrals of
motion is formulated. From the plot of the Lyapunov exponent we see
that analytical formulae~\eqref{boundary_chaos} give an accurate
definitions of the chaotic zone boundaries. Note that the
afore-mentioned technique can also be applied to the pure undulator
field (there is no the axial field). In this case from
Eq.~\eqref{chaos_condit} it follows a simple chaotization condition
that in terms of paper~\cite{ChenDavidson1990} reads
\begin{equation}\label{chaos_pure_undul}
  a_w \geq \beta_{zb} \gamma_b/(\sqrt{2}\cosh[k_w x_b]).
\end{equation}
In Fig.~\ref{fig_chaos_undul} the regular and chaotic regions in the
parameter space $(k_w x_b,a_w)$, according to
Eq.~\eqref{chaos_pure_undul}, are shown. Comparing Fig.~2
of~\cite{ChenDavidson1990} and Fig.~\ref{fig_chaos_undul} of the
present paper we notice that the proposed simple estimation is in a
reasonable agreement with the numerical simulation result cited in
paper~\cite{ChenDavidson1990}.

Using the information about the electron's dynamics in the pump
magnetic field one can subsequently analyze the dynamics of ensemble
of electrons in view of the ponderomotive wave. In a special
coordinate system (that moves on the invariant torus surface if
there is no the ponderomotive wave) one can split degrees of freedom
and partially integrate the equations of
motion~\cite{Zaslavskii_1988}. As a result, the universal reduced
model of the FEL~\eqref{reduced_model} that incorporates the
intricate dynamics of electrons and the intramode scattering is
derived.

What is important is that there are two types of detuning in reduce
model~\eqref{reduced_model}: `axial' detuning that changes along the
interaction region via the profiled waveguide width and the
`transverse' detuning that changes across the beam because the pump
magnetic field is inhomogeneous and the average velocity of the
electron depends on its initial transverse position. The
"transverse" detuning causes the thick beam layering and the
degradation of the external layers' contribution into the total
efficiency (the degradation is particulary strong near the
magnetoresonance). In the present paper I demonstrate that this
problem and the saturation effect can be overcome by the control of
the beam-wave interaction. The physical mechanism of such a control
is that by changing the waveguide profile one control the axial
detuning and thus regulate the phase shift between the ponderomotive
wave and current. This phase shift defines the transfer of the
energy between the beam and the wave and its regulation allows one
to optimize the interaction.

The practical example of optimization of the FEL, whose parameters
are close to those of the experiment~\cite{Destler_1996}, is
demonstrated. The simulation results based on the non-simplified
model (see Sec. II) strongly indicate that combining the
magnetoresonance effect with the optimized profile waveguide one can
enhance the FEL efficiency by a factor of five or six. The
efficiency in the experiment~\cite{Destler_1996} was around 4\%.
Applying the axial magnetic field the efficiency has been  increased
up to nearly 12\%, but about 30\% of electrons do not interact with
the wave because of the initial transverse detuning. Following the
waveguide optimization the efficiency has reached 22\%, in
particular, due to a much more effective interaction between the
external beam layers and the wave.

\appendix

\section{Time-dependent excitation equation of an irregular waveguide}
\label{appendix1}

The evolution of the resonant (synchronous) TE$_{01}$ mode is
governed by the $x$-component of the vector-potential $A_x^s$, which
satisfies the wave equation
\begin{equation}\label{wave_eq_Ay}
g \Bigl(\nabla^2 - \frac{1}{c^2}\frac{\partial^2}{\partial t^2}\Bigr)A_x = -\frac{4\pi}{c} j_x.
\end{equation}
We seek a solution to the equation of the form:
\begin{equation}\label{Fourier_amplitude}
  A_x(\vec r, t) =\mathrm{Re}\int_{0}^\infty  \bar V_(z,\omega')\sqrt{\frac{b}{w(z)}}
        \cos\Bigl(\frac{\pi y}{w(z)}\Bigr) \,e^{-i\omega' t}\,  d\omega'.
\end{equation}
Substituting \eqref{Fourier_amplitude} into \eqref{wave_eq_Ay} we
derive the excitation equation for the Fourier amplitude $\bar
V_(z,\omega')$
\begin{multline}\label{nonregular_amplitude}
  \Bigl\{\frac{\partial^2}{\partial z^2} + k_z^2(z,\omega')\Bigl\}\bar V_(z,\omega')=
                  -\frac{8\pi}{Sc} \sqrt{\frac{b}{w(z)}} \int_{-a/2}^{a/2}  dx
                  \times
  \\
                  \int_{-w/2}^{w/2} dy \, \bar j_x(\vec r,\omega')  \cos\Bigl(\frac{\pi y}{w(z)}\Bigr),
\end{multline}
where $ k_z^2(z)= (\omega/c)^2 - (\pi/w)^2 - (w'/2w)^2(1+\pi^2/3)$
and $ \bar j_x(\vec r,\omega') = \pi^{-1}\int_{-\infty}^{\infty}
j_x(\vec r,t)e^{i\omega' t} dt$. Here $\bar j_x(\vec r,\omega')$ is
the Fourier amplitude of the current density. We will consider that
at the section $z=0$ the FEL-amplifier is seeded by the TE$_{01}$
mode with a frequency of $\omega$ and amplitude $V_0$, and the
interaction region is ideally matched to a regular output waveguide
at the section $z=L$
\begin{equation}\label{boundary_condit}
\begin{split}
  & \Bigl(\frac{\partial \bar V}{\partial z} + ik_z \bar V\Bigr)\Bigr|_{z=0} = 2ik_z V_0\delta[\omega'-\omega],
  \\
  & \Bigl(\frac{\partial \bar V}{\partial z} - ik_z \bar V\Bigr)\Bigr|_{z=L} = 0.
\end{split}
\end{equation}
The conditions for the waveguide profile at the ends of the
interaction region have the following form: $w'(0)=w'(L)=0$. We
assume that $\bar j_x(\vec r,\omega')$ is the narrow-band signal
with a fundamental frequency of $\omega$. This means that the
current density can be written as $
  j_x(\vec r, t) = \mathrm{Re}\{J_x(\vec r,  t)e^{-i\omega t}  \},
$
where $J_x(\vec r, t)$ is a slow function of time such that
\begin{multline*}
  J_x(\vec r, t) \approx \frac{\omega}{\pi}\int_{t-\pi/\omega}^{t+\pi/\omega} j_x(\vec r,t)e^{i\omega t} dt
  \approx
\\
   \int_{-\infty}^{\infty} \bar j_x(\vec r,\omega+ \Delta \omega)
                         e^{-i\Delta  \omega t}d(\Delta\omega), \quad \Delta\omega = \omega'-\omega.
\end{multline*}
Expanding $k_z^2(z,\omega')$ into Taylor's series over $\omega$ up
to the linear term, multiplying the Eq.~\eqref{nonregular_amplitude}
by $e^{-i\Delta \omega t}$ and integrating it over $\Delta\omega$
from $-\infty$ to $\infty$ we derive the time-dependent excitation
equation~\eqref{nonregular_amplitude_time} for the slow in time
amplitude $V(z,t) = \int_{-\infty}^{\infty} \bar V(z,\omega+ \Delta
\omega) e^{-i\Delta  \omega t}d(\Delta\omega)$. The
solution~\eqref{Fourier_amplitude} and the boundary
conditions~\eqref{boundary_condit} can be rewritten as
Eq.~\eqref{microwave} and Eq.~\eqref{boundary_condit_time},
respectively.

\section{Intramode scattering in an irregular waveguide}
\label{appendix2}

The Eq.~\eqref{nonregular_amplitude_time} is used for the numerical
simulation of non-averaged model of the FEL, but for the analytical
study we have to rederive  the excitation equation in a different
form. We seek a solution to Eq.~\eqref{nonregular_amplitude} for
microwave Fourier amplitude $\bar V$ of the form:
\begin{equation}\label{wave_amplitude_split}
  \bar V = \Bigl(\bar V_+ e^{i\Psi} + \bar V_- e^{-i\Psi}\Bigr)\Bigl(\frac{k_z^0}{k_z}\Bigr)^{1/2},
            \; \Psi(z) = \int_0^z k_z(z')dz',
\end{equation}
($k_z^0\equiv k_z(0)$) and we allow amplitudes $\bar V_\pm$ to be
functions of the axial position $z$.  Applying the method of
variation of constants we derive first-order equations for new
unknown functions $\bar V_\pm(z)$
\begin{subequations}\label{Fourier_amplitude_split}
\begin{align}
\label{Fourier_amplitude_split1}
 \frac{d\bar V_+}{dz} & = -\frac{i\, RHS\,e^{- i\Psi}}{2\sqrt{k_z^0k_z}}+
                            \frac{\partial k_z}{\partial z}\frac{\bar V_-\,e^{- 2i\Psi}}{2k_z}
  \\
\label{Fourier_amplitude_split2}
  \frac{d\bar V_-}{dz} & = \frac{i\, RHS\, e^{i\Psi}}{2\sqrt{k_z^0k_z}}+
                            \frac{\partial k_z}{\partial z}\frac{\bar V_+\,e^{2i\Psi}}{2k_z}.
\end{align}
\end{subequations}
Here $RHS(z)$ is the right-hand side of
Eq.~\eqref{nonregular_amplitude} and the boundary conditions become
\begin{equation}\label{boundary_condit_split}
  \bar V_+\bigr|_{z=0} = V_0\delta[\omega'-\omega],
   \quad \bar V_-\bigr|_{z=L} = 0.
\end{equation}
Note that Eqs.~\eqref{wave_amplitude_split},
\eqref{Fourier_amplitude_split},  \eqref{boundary_condit_split}
formally define the exact solution to
Eq.~\eqref{nonregular_amplitude}. Now restrict ourself to the case
of the resonant interaction of a beam with a forward microwave. The
first term in the right-hand side of
\eqref{Fourier_amplitude_split1} describes the above-mentioned
resonant interaction and should be taken into account, whereas the
first term in the right-hand side of
\eqref{Fourier_amplitude_split2} is nonresonant and might be
omitted. We will also assume that $k_z^{-1}\partial_z k_z < 1$ and
neglect the second term in the right-hand side of
\eqref{Fourier_amplitude_split1} because it describes the
second-order scattering effect (according to the boundary conditions
for a source-free regular waveguide $\bar V_-(z)\equiv 0$), but we
will keep the second term in \eqref{Fourier_amplitude_split2}. Using
some algebra and performing the inverse Fourier transformation we
derive Eqs.~\eqref{microwave_split} and \eqref{time_amplitude_split}
for $V_+(z,t) = \int_{-\infty}^{\infty} \bar V_+(z,\omega+ \Delta
\omega) e^{-i\Delta  \omega t+i\Delta
\omega\partial_\omega\Psi}d(\Delta\omega)$ and $V_-(z,t) =
e^{-2i\Psi}\int_{-\infty}^{\infty} \bar V_-(z,\omega+ \Delta \omega)
e^{-i\Delta \omega t -i\Delta
\omega\partial_\omega\Psi}d(\Delta\omega)$.

\bibliography{goryashko}

\end{document}